# Coherent Virtual Absorption in Dielectric Slabs: A Temporal Analysis of Symmetric and Asymmetric Geometries

Kaizad Rustomji[1], Nasim Mohammadi Estakhri[1,2], Nooshin M. Estakhri[1,2,3]*

[1] *Fowler School of Engineering, Chapman University, Orange, California 92866, USA*

[2] *Schmid College of Science and Technology, Chapman University, Orange, California 92866, USA*

[3] *Institute for Quantum Studies, Chapman University, Orange, California 92866, USA*

*nmestakhri@chapman.edu

**Abstract:** Coherent virtual absorption refers to time-limited storage of optical energy in lossless configurations due to excitation of a complex zero frequency through proper temporal engineering of the incident wave. Given the dynamics underlying the effect and the storage-release mechanism occurring for finite excitation pulses, studying and understanding the associated time dynamics are crucial for enabling future applications. In this work, we carefully investigate this phenomenon in symmetric and asymmetric geometries, shedding light on practical considerations in situations when a closed-form analytical solution is not readily available. Combinations of time domain analysis and spectral filtering are used to enable systematic analysis of these structures. Our approach can be generalized to more complex structures, including multilayered and inhomogeneous cases, providing new opportunities for optimized energy storage and advanced sensing applications utilizing complex-frequency dynamics in lossless designs.

# Introduction

Across physics, from high-energy particle phenomena to electromagnetics, the scattering matrix (S-matrix) is arguably the most widely used framework, encapsulating the input–output response of a system. Mathematically, the S-matrix lives in the complex vector space, and studying electromagnetic systems near its poles and zeros has become particularly compelling, as the ability to engineer their locations and selectively excite them can lead to interesting physics. For instance, in a linear scattering system, coherent perfect absorption can be achieved by shifting one of the S-matrix zeros onto the real axis through engineering the loss of the system [1–4]. On the other hand, gain can be used to shift an S-matrix pole to the real axis, marking the lasing threshold [5,6], and a pole–zero convergence on the real axis produces a bound state in the continuum [7,8]. Driven by the exciting science and potential applications of accessing and systematically manipulating these singularities, complex frequency excitations have recently emerged as a novel platform. This field seeks to exploit and engineer the behavior of systems under nontrivial temporal conditions, where singular properties of the system are not accessible through real-frequency excitations. [9–12]. In an ideal form, complex-frequency excitation signals are those of the form $e^{-i\omega t}$ with a complex frequency $\omega = \omega' + i\omega''$; where $\omega'$ sets the oscillation frequency and $\omega''$ parameter governs exponential growth or decay, so the term "complex" also reflects that the excitation combines both oscillatory and exponential behavior. Exploiting such transient waveforms, one can mimic virtual gain and loss and possibly access the interesting physics that lie at the singularities of the S-matrices in the complex vector space, through temporal engineering of the excitation signal [13–17].

In this paper, we focus on complex frequency excitation for the purpose of virtual absorption and, in particular, coherent virtual absorption. Virtual absorption refers to mechanisms where variations in the system or incident signal(s) create a virtual loss effect in a lossless structure where either reflection or transmission vanishes [18,19]. In a coherent virtual absorption effect, a lossless structure under illumination exhibits zero scattering and temporally stores the entire incident energy. For instance, the system might temporarily store the incoming energy, which is specifically tailored to mimic a complex frequency excitation, until the input wave is switched off. Such phenomena may be achieved in electromagnetics and optics [6,20,21], elastodynamics [22] or acoustics [23,24]. Using a combination of finite-difference time-domain (FDTD) and direction-resolved spectral filtering, here we investigate scattering properties of multilayered dielectric slabs, demonstrating highly asymmetric coherent control of absorption and scattering through complex excitations. Our findings, together with the discussions on the intricacies of different computational approaches for these problems, enable the broader exploration of more complex structures, such as inhomogeneous, reconfigurable, multilayered metasurfaces and complex scattering configurations [25–32], where additional degrees of freedom can be systematically optimized.

## Results and Discussion

We start by considering a single-layer slab as shown in Figure 1(a). Under normal illuminations from both sides, the slab creates a 2-port system whose S-matrix can be readily calculated analytically [33]. For S-matrix calculations, the input waves are set to have a complex frequency $\omega = \omega' + i\omega''$, and are polarized along the y-axis. Using this framework, the eigenvalues of the S-matrix are then calculated in the complex domain, resulting in an infinite set of paired zeros and poles, associated with complete absorption or lasing conditions, respectively [4,18,34,35]. Here, we focus on the zeros of the eigenvalues to induce and temporally observe the coherent virtual absorption effect in lossless geometries. In this framework, the eigenvector corresponding to each eigenvalue determines the relative amplitude and phase of the needed excitations at the corresponding ports to induce the effect. If the two ports of the system are placed symmetrically relative to the center of the slab, eigenvector components are inevitably equal in amplitude, and the relative phases of the two incident beams need to be set at 0 or 180 degrees, set by the zero eigenvalue undergoing excitation. Considering the $e^{-i\omega t}$ time convention in the temporal analysis, this effect is associated with a certain set of complex frequencies in the upper half-plane.

In the first step, a complex zero of the system for a slab with a refractive index $n = 3$, as shown in Figure 1(a), is identified. As mentioned, the system possesses an infinite set of zeros from which we choose the point matching $kL = 4.189 + i0.231$ in which $k$ is the complex wavenumber associated with $\omega = \omega' + i\omega''$. The slab is then illuminated from both sides with plane waves of temporal dependence $E_{inc}(t) = E_0 e^{\omega'' t} e^{-i\omega' t} u(-t) + E_0 e^{-(t/\tau)^2} e^{-i\omega' t} u(t)$.

The exponential growth for $t < 0$ signifies the complex frequency excitation, determined by the imaginary part of the frequency $\omega''$ and the Gaussian decay for $t \geq 0$ provides a smooth cutoff after the pulse is switched off. Choosing $\tau = 1/\omega'$ will ensure that the pulse is effectively switched off within a single oscillation. We define $t = t_0$ as the moment when the maxima of the incident pulse envelopes reach the edges of the slab located at $x = \pm L/2$, marking the time when the specific complex-frequency portion of the incident pulse has fully undergone interaction with the slab.

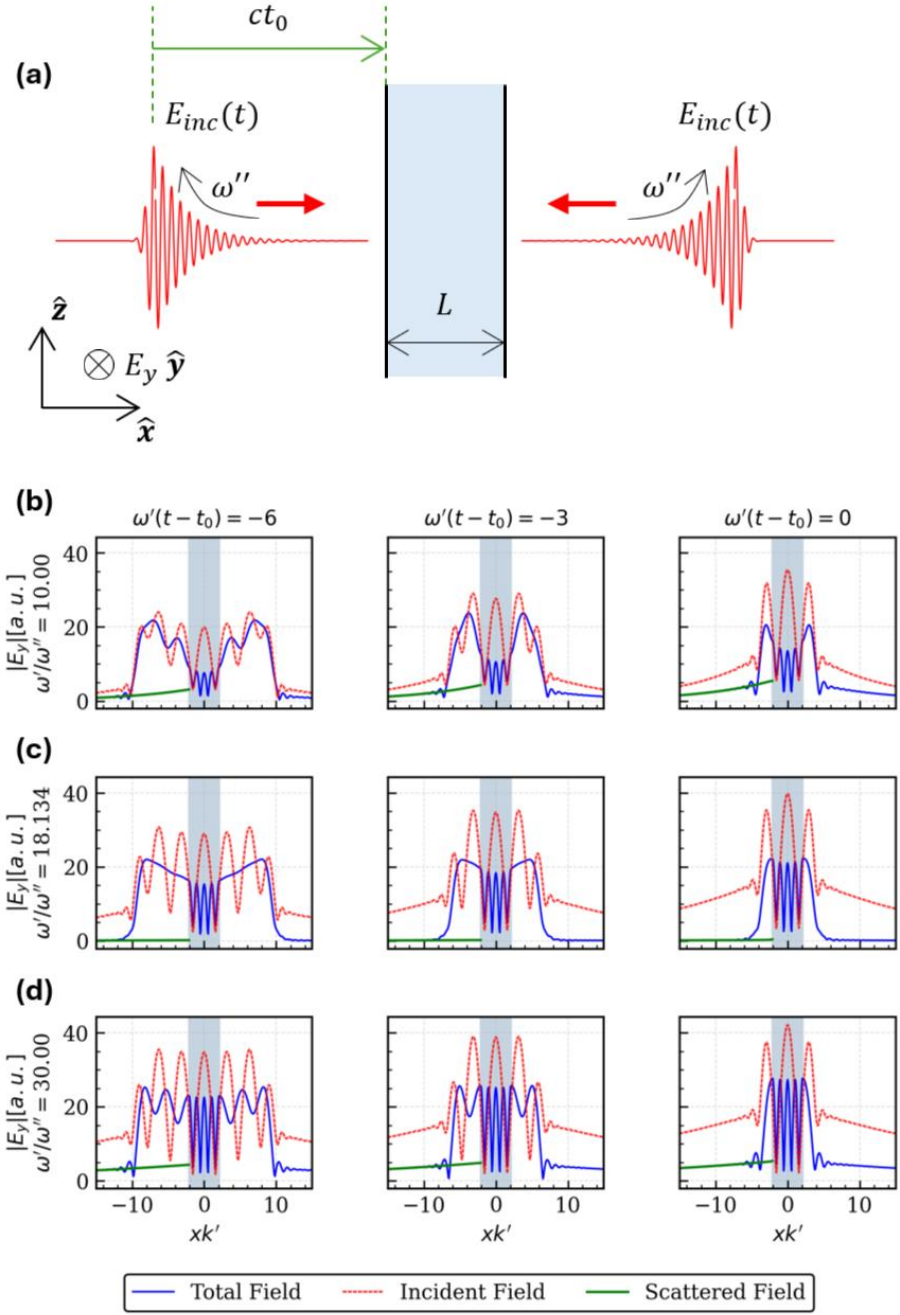

**Figure 1** (a) Schematic illustration of the dielectric slab illuminated by two counterpropagating pulses. The refractive index of the slab is set to $n = 3$ and the structure is infinitely extended in the z direction. (b)-(d) time evolution of the total field (blue), incident field (dashed, red), and the scattered field (green) for different values of $\omega'/\omega''$. For all cases, $\omega'$ is fixed at the location of zero and $\omega''$ is varied. The shaded blue region shows the location of the slab.

Operating at the complex frequency associated with a zero eigenvalue of the system implies that the slab should not scatter the incident energy if the illumination is sustained. However, once the input signal is switched off in practical settings, non-zero scattering is expected to occur. On this account, a temporal study of the scattering response is necessary to observe such transient scattering signals. While analytical solutions can be constructed for a one-dimensional slab under planewave excitations [16], more complex geometrical configurations, which are frequently adopted in practical and experimental contexts, must rely on computational approaches to analyze the scattering response. With this goal in mind, here, we use an FDTD technique to capture different signatures of virtual absorption both in the time and frequency domains. For our specific choice of complex zero, the coherent virtual absorption occurs at $\omega'/\omega'' \approx 18.134$ and the two incident waves must be in-phase. We can modify the induced virtual loss through modification of $\omega''$ by moving up or down in the complex plane. This is shown in Figures 1(b)-(d) where the temporal evolution of the total electric field around the slab is plotted (blue curves). As the incident pulse propagates into the slab, the total field inside the dielectric region gradually increases in time. For the case of the zero of scattering eigenvalue in Figure 1(c), the amplitude of the total electric field is smooth with no ripples, indicating that no scattering (from either side) is present prior to the transition point $t = t_0$. This is true while at other complex-frequency excitation points (Figure 1(b) and (d)) the ripples in the total electric field curves clearly indicate the presence of scattering from this lossless geometry. Here, we also calculate the one-sided scattered field (green curves) as the difference between the total field and the incident field from the left-hand side. Due to partial reflection and transmission of the two incident waves, the scattered field is non-zero for complex frequencies with a nonzero Scattering matrix eigenvalue, while at the virtual absorption frequency, the scattering is strongly suppressed, as expected.

Using a time domain analysis technique, the transient behavior and dynamics of the waves as they interact with the structure (before and after switching) are readily available. To better understand the temporal signatures of the coherent virtual absorption effect here, we next look at the buildup and release of the energy in the system. To this end, we place a field monitor outside of the left edge of the slab at a subwavelength distance of $2/k'$, with the location of the monitor denoted by $x_{monitor}$. We also define $t_{monitor}$ as the time when the peak of the left incident pulse reaches the observation point at $x_{monitor}$ after initially reaching the edge of the slab, then reflecting to reach the location of the monitor. Therefore, in an ideal case, at the virtual absorption point, the measured instantaneous scattered power at the location of the monitor, $x_{monitor}$ is expected to be exactly zero up until $t = t_{monitor}$.

To probe the numerical behavior of this phenomenon, the calculated total instantaneous power defined as $Re(E_y(x_{monitor}, t))Re(H_z(x_{monitor}, t))$ at the observation point, when operating at the virtual absorption point, is plotted in Figure 2(a). As can be seen, the total instantaneous power is initially positive since the incident wave in this region is traveling in the +x direction without any reflections or any transmissions from the right-hand side. The instantaneous power, however, reverses sign at $t \approx t_{monitor}$, marking the onset of scattering when the switched off exponential growth induces nonzero scatterings. We can also temporally or spectrally isolate the *scattered* signal. In the following, we first consider the temporal results and will discuss the spectral isolation technique and its results in the next section. The temporal isolation is performed by subtracting a free-space reference fields $E^{inc}$ and $H^{inc}$ propagating from left to right and generated through a comparable FDTD simulation in the absence of the slab from the total electric field in the presence of the slab. Using these scattered electric and magnetic fields, the evolution of the scattered energy in time is calculated and shown in Figure 2(b). Interestingly, a region of fast increase in scattered energy is observed upon switching off the exponential growth, which is followed by gradual accumulation of scattering and eventually flattening out, corresponding to the finite length of the signal. Similarly, no scattering occurs until $t = t_{monitor}$ and if the numerical monitor is placed closer to the slab, the scattered energy rise point will get closer to $t_0$.

Looking at Figure 2(b), the energy is not scattered until $t = t_{monitor}$, and given that the slab and the surrounding environment are lossless, this implies that the energy is transiently stored within the structure. To quantify this effect, we also plot the evolution of the stored energy inside the slab in Figure 2(c), which reaches its maximum at $t = t_0$ before decaying, as expected. Figures 2(a)-(c) clearly demonstrate the virtual absorption/storage of the pulse inside the slab at the complex frequency associated with the zero of the scattering eigenvalue. To quantify the scattering energy that occurs at complex frequencies not matching zero points, we also plot the scattered energy for various complex frequencies by modifying the imaginary part of the complex excitation frequency $\omega''$ and traversing the complex plane upward and downward through the virtual absorption point, as shown in Figure 2(d). Here, we are plotting the normalized scattered energy defined as total scattered energy at $t = t_{monitor}$ divided by the total energy of the incident pulses.

In agreement with the theoretical predictions, the scattered energy is at its lowest value at the virtual absorption point happening at $\omega'/\omega'' = 18.134$. We note that in our simulations, the pulse shutoff time $\tau = 1/\omega'$ can also impact the onset of scattering, as studied in the inset of Figure 2(b). Indeed, reducing $\tau$ to $0.2/\omega'$ will push the scattering start time closer to $t = t_{monitor}$, as expected.

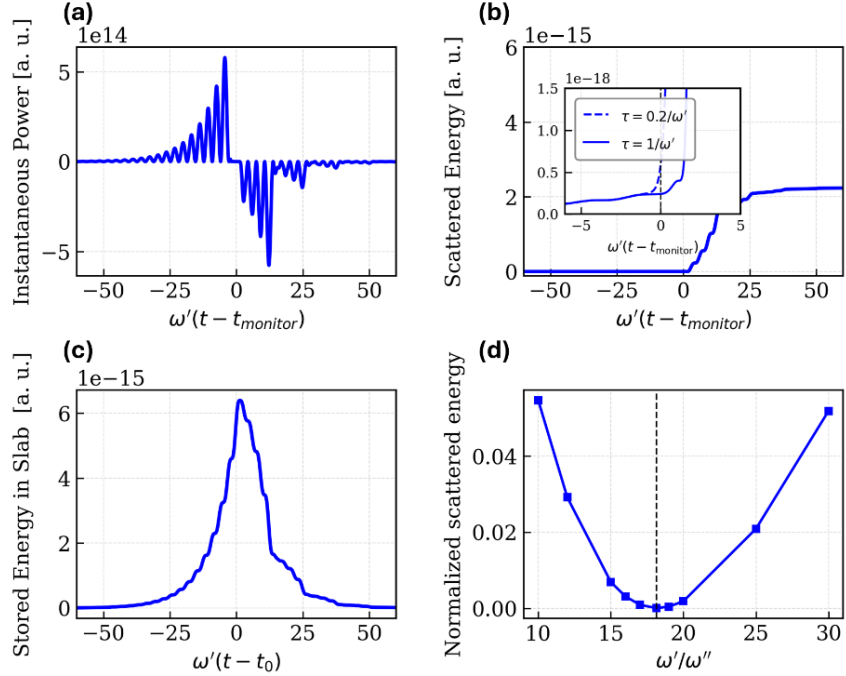

**Figure 2** (a) Time evolution of the instantaneous power at the observation point $x_{monitor}$ located at distance of $2/k'$ from the boundary of the slab, at the complex frequency associated with coherent virtual absorption (b) scattered energy to the left, and (c) total energy stored in the slab. In both panels the dashed lines show the results for complex excitation at the on-zero eigenvalue point of $kL = 4.189 + i0.231$. Inset in panel (b) shows a zoomed-in version for slow (solid) and fast (dashed) switch times. (d) Normalized scattered energy at $t = t_{monitor}$ for different values of $\omega'/\omega''$.

As indicated earlier, besides fully temporal techniques, the scattered signal may also be isolated through a spectro-temporal decomposition approach. We note that while temporal response is more readily available through the FDTD approach or similar time-domain simulation techniques, direct calculations of the transient *scattered field* in a general setup, such as multiport or multimode configurations, can quickly become complicated. To circumvent these limitations, next, we utilize an alternative method based on spectral filtering to efficiently isolate the scattered signal. Essentially, in this scheme, the temporal signal is first transformed into the spectral domain, specific spatial frequency components are then isolated and subsequently reconstructed in the time domain, enabling us to effectively extract the scattered fields in both time and spatial frequency domains. In the slab configuration, we achieve this by monitoring the fields across the $z = 0$ line, extending between the position of the source and the left edge of the slab. Figure 3(a) shows the spatiotemporal evolution of the electric field $E_y(x,t)$ corresponding to the excitation at the virtual absorption complex frequency. As time progresses, the wave initially propagates unidirectionally from left to right till $t = t_0$ at which it gets scattered by the slab and partially reflected, propagating in the opposite direction ; i.e., $e^{+ik'x} \rightarrow e^{-ik'x}$. To isolate the contribution of scattering fields, we apply a space-time Fourier transform on the electric field $E_y(x,t)$ yielding the spatiotemporal spectrum $E_y(k,\omega)$ as shown in Figure 3(b). The spectra exhibit hotspots at $k = \pm k_0$ and $\omega = \pm \omega'$, corresponding to the right-propagating incident wave and the left-propagating scattered wave. The negative temporal frequencies naturally arise as the spectrum is calculated from real-valued time-dependent electric field $Re(E_y(x,t))$.

Transformation into the spectral domain allows us to separate the scattered field from the incident field in the context of coherent virtual absorption phenomenon. Since the scattered components travel with negative $k$-values, we can effectively isolate them by applying a mask in the $(k,\omega)$ space as shown in Figure 3(b). The masking is carried out by element-wise multiplication in the spatiotemporal frequency space. An inverse Fourier transform of the masked spectrum (Figure 3(c)) yields the isolated scattered field $E_y^{scat}(x,t)$ shown in Figure 3(d) for our example. We note that when operating in a multimode structure, the mask can be modified to isolate the scattering of the desired mode.

Taking a slice of Figures 3(a) and 3(d) at the position of the monitor $x = x_{monitor}$, placed outside of the slab, we also look at the temporal evolution of the total field in Figure 3(e) and the spectrally filtered scattered field shown in Figure 3(f). As expected, spectral filtering eliminates the incident portion of the total field, allowing the scattered field to be effectively isolated and compared with the previous approach in Figure 4(a). As we are operating at the scattering zero of the system, the release of energy occurs only after $t_{monitor}$ (see Figures 3(f) and 4(a)). Compared to the temporal approach, this approach allows us to readily separate and isolate spatiotemporally overlapping scattering modes (i.e., modes carrying different momenta) in the $(k,\omega)$ space for multimode or multiport configurations. We also perform a similar spectral filtration to obtain the scattered magnetic field $H_z^{scat}(x,t)$, and compute scattered power and energy. The results are shown in Figure 4(b) for both temporal and spectral approaches. We note that the spectral method evaluates the scattered power based on a limited-capacity time/space monitor. As such, it may slightly underestimate the total scattered energy when the structure is excited at off-zero conditions. Nevertheless, our simulations confirm that the outcomes of the two methods match very well.

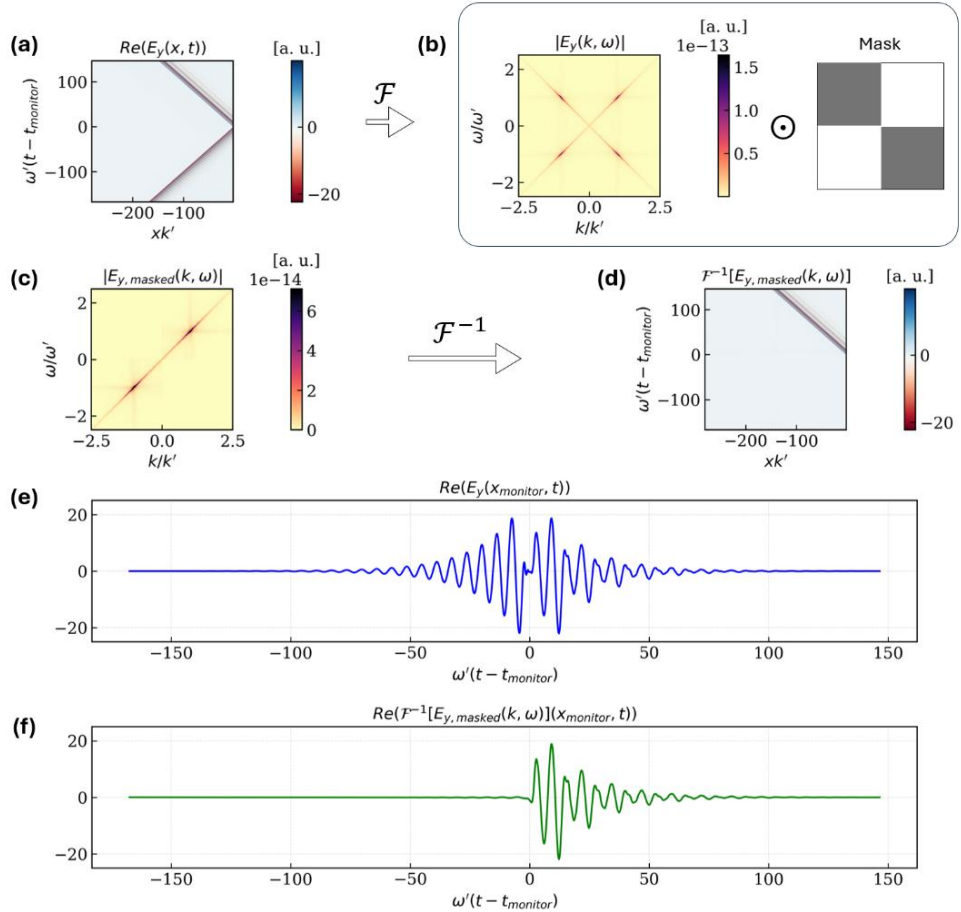

**Figure 3** (a) Spatiotemporal evolution of the electric field, $Re(E_y(x,t))$, monitored on a horizontal line on the left side of the slab. (b) Space-time Fourier transform of the electric $E_y(k,\omega)$ and the filtering mask (c) Electric field in the Spatiotemporal Fourier domain after the application of the mask, $E_{y,masked}(k,\omega)$ (d) Spatiotemporal evolution of the scattered electric field obtained from $E_{y,masked}(k,\omega)$. (e) Total electric field at the position of the point monitor $x_{monitor} = -L/2 - 2/k'$ (f) Scattered portion of the electric field at the location of the point monitor.

Finally, we also use the spectral analysis approach to further investigate the dispersion behavior of scattering from the slab when the complex excitation frequency is slightly deviated from the virtual absorption point (similar to Figure 2(d)). Results are presented in Figures 4(c) and 4(d), where the imaginary and real parts of the complex excitation frequency are changed around the coherent virtual absorption point, respectively. Expectedly, both approaches consistently show that the scattered energy vanishes at the virtual absorption complex frequency. We have also numerically calculated and plotted the same quantity for a standard Gaussian incident pulse with a temporal profile of $E_{inc}(t) = E_0 e^{-(t/\tau)^2} e^{-i\omega' t}$ where $\tau = 1/\omega''$. As expected, the scattered energy, while small, cannot completely vanish. We note that the central frequency of the Gaussian pulse is deliberately selected to be equal to the real part of the corresponding complex-frequency excitations to emphasize the key role of a tailored loss in achieving coherent virtual absorption through a complex-frequency excitation.

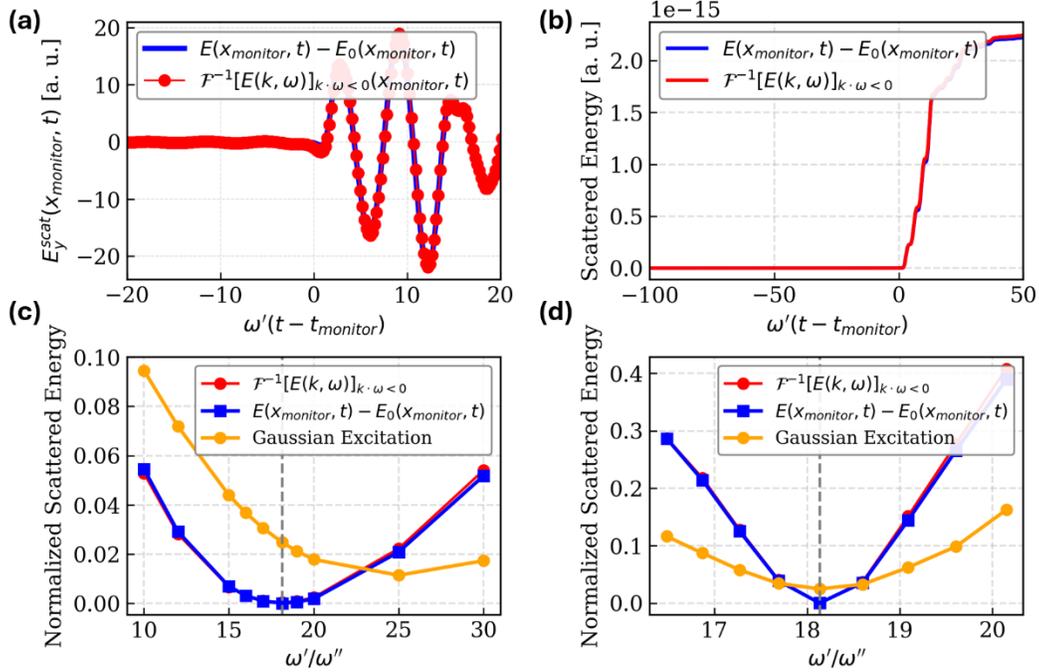

**Figure 4** (a) Scattered portion of the electric field at the location of the point monitor. Blue line indicates values directly calculated via FDTD simulation and red circles denote the results from the spectral filtering method. (b) Scattered energy calculated at $x_{monitor}$. (c) Normalized scattered energy at $t = t_{monitor}$ for different values of $\omega'/\omega''$, varying $\omega''$ or (d) $\omega'$ around the virtual absorption frequency. The orange line indicates the same quantity calculated for a standard Gaussian pulse.

The dielectric slab studied above is geometrically symmetric. Consequently, coherent control on the scattering and/or absorption properties of such a symmetric system inevitably requires a symmetric excitation from all ports. Breaking such symmetry, on the other hand, will enable us to potentially control and regulate the response to an excitation from one side of the slab with a much weaker *control* signal applied from the other side. With this goal in mind, we next investigate virtual absorption in an asymmetric two-layer configuration as schematically shown in Figure 5(a). The structure consists of two conjoined slabs with refractive indices $n = \sqrt{2}$ (left side) and $n = 3$

(right side), each with equal thickness of $6.93/2k'$. This system also possesses an infinite set of eigenvalue zeros in the complex plane from which we choose $kL = 6.93 + i72$ in which $k$ is the complex wavenumber associated with $\omega = \omega' + i\omega''$. The multilayer structure is illuminated from both sides with plane waves of temporal dependence $E_{inc}(t) = E_0 e^{\omega'' t} e^{-i\omega' t} u(-t) + E_0 e^{-(t/\tau)^2} e^{-i\omega' t} u(t)$, similar to the previous scenario. The scattering zeros are calculated analytically here, and it is worth noting that the following discussions are applicable to any similar multilayer structure. While we investigate a sample two-layer structure here, the degrees of freedom available in a general multilayer structure may be utilized to position the scattering zero at a desired frequency and to arbitrarily control the ratio between the excitation beams. Considering the structure described above, the eigenvector of the system at the chosen zero point is calculated, resulting in a relative phase of 160.66° and an unbalanced power ratio of $P_{inc}^{(R)}/P_{inc}^{(L)} \approx 0.105$ between the two excitation pulses. In other words, the incident field illuminating the structure from the right side is now carrying approximately 90% less power than the beam illuminating the structure from the left-hand side.

Using FDTD to capture the temporal signatures of virtual absorption, we investigate the evolution of the electric field as it interacts with the multi-layer structure, as shown in Figures 5(b)-(d). For our choice of complex zero, the coherent virtual absorption occurs at $\omega'/\omega'' = 9.625$ and in all the plots the same relative phase and amplitude are assumed. Simulations confirm that scattering is suppressed at the calculated complex frequency (i.e., when $\omega'/\omega'' = 9.625$), as expected. As the incident pulse propagates into the slab, the total field inside the dielectric layers gradually increases in time. For the case of the zero of the scattering eigenvalues in Figure 5(c), the amplitude of the total electric field is smooth with no ripples outside the multilayer geometry, indicating that no scattering (from either side) is present until time $t = t_0$. In contrast, for the cases of non-zero eigenvalue shown in Figures 5(b) and 5(d), scattering signatures are evident, appearing as ripples in the total electric field. Consistent with our previous studies, the one-sided scattered fields are also calculated as the difference between the total field and the incident field from the left/right and shown with green curves. Specifically, the scattered field is non-zero for the complex frequencies deviating from the virtual absorption point while it is strongly suppressed at the design point corresponding to zero scattering eigenvalues, as expected.

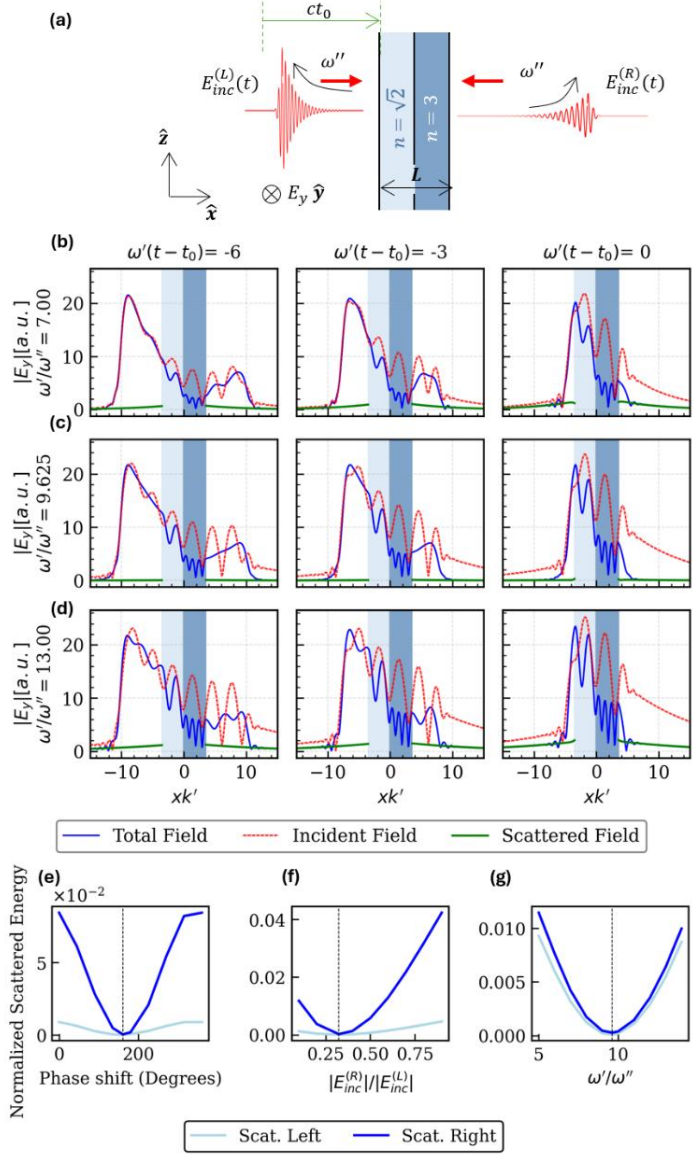

**Figure 5** (a) Schematic illustration of the two-layer geometry illuminated by two counterpropagating pulses. The refractive indices of the right and left slabs are set to $n = 3$ and $n = \sqrt{2}$, respectively, and the structure is infinitely extended along the z-direction. (b)-(d) Time evolution of the total field (blue), incident field (dashed red), and the scattered field (green) for different values of $\omega'/\omega''$. For all cases, $\omega'$ is fixed at the location of zero and $\omega''$ is varied. The shaded blue regions show the location of the two layers, showing clear signatures of coherent virtual absorption in the multilayer geometry (e)-(g) Normalized scattered energy till $t = t_{monitor}$ or (e) different values of phase shift between excitation signals, (f) different relative amplitudes of the excitation signals (varying $|E_{inc}^{(R)}|$ ), and (g) different values of $\omega'/\omega''$ by varying $\omega''$ in the vicinity of the virtual absorption frequency. The dark and light blue curves correspond to the numerically calculated normalized scattered energy to the right and left side of the structure, respectively. The dashed vertical lines in (e)-(g) showed the design coherent virtual absorption point.

We also investigate the behavior and sensitivity of the scattering response to the system input variables by systematically analyzing the effect of varying excitation parameters. We use the spectral analysis approach described above to calculate the normalized scattered energy at the location of two point-monitors placed at $x_{monitor}^{(L)}$ and $x_{monitor}^{(R)}$, located at $2/k'$ from the left and right edges of the two-layer configuration, respectively. Figures 5(e)-(g) illustrate the normalized scattered energy at $t = t_{monitor}$, calculated at each side of the structure as a function of the relative phase between the two excitation pulses, their amplitude ratio, and variations in the complex excitation frequency, which are derived while always holding the other two parameters fixed. The results confirm the presence of the coherent virtual absorption for the exact theoretically calculated conditions, indicated by the dashed vertical lines in all plots. As we expect, all cases exhibit a clear increase in scattered energy from the multilayer lossless structure as the conditions deviate from the ideal point.

## Conclusion

In this work, we use full-wave time-domain simulations to perform a comprehensive temporal analysis of coherent virtual absorption in symmetric asymmetric planar geometries. We systematically analyze signatures of coherent virtual absorption phenomena, and the implications of operating in the vicinity of the zero-scattering complex frequency. In addition to intuitive time-domain measures, we also present a method to quantify scattered fields in the spectral domain which shows excellent accuracy and could be used for analysis of multiport and multimode systems. We demonstrate that virtual absorption can be generalized to asymmetric excitation amplitudes, which open new degrees of freedom for coherent control of absorption and scattering. We investigate the sensitivity of virtual absorption to excitation parameters including frequency, phase, and amplitudes, offering opportunities for dynamic wave control and sensing applications [36]. We envision that inverse optimization approaches [37–41] could be valuable to efficiently tailor complex excitations in engineered metamaterials [42] and plasmonic structures [43], opening new avenues for exploiting complex excitations toward tailored energy storage and dynamic scattering control.

## Methods

FDTD simulations are performed using the Flexcompute Tidy3D simulation software [44],with a free-space central wavelength of $\lambda_0 = 600$ nm. A uniform mesh size of $\lambda/80$ is used in all simulations. The sources are two counterpropagating plane waves with the electric field polarized along the *y*-axis, propagating in the *x*-direction. The sources are placed symmetrically at $x = \pm 280/k'$, normally incident on a centrally located slab. Perfectly matched layers are used as boundary conditions in the *x*-direction and Bloch periodic boundary conditions are applied in the *z*-direction. The Courant factor in all simulations is set to 0.99, which leads to a time step of $1.75 \times 10^{-5}$ ps, the maximum duration of simulation was set to be 3 ps or until the instantaneous integrated E-field intensity dropped below $10^{-5}$, whichever occurred first.

## References

1. L. Chen, T. Kottos, and S. M. Anlage, "Perfect absorption in complex scattering systems with or without hidden symmetries," Nat. Commun. **11**, 5826 (2020).

2. D. G. Baranov, A. Krasnok, T. Shegai, A. Alù, and Y. Chong, "Coherent perfect absorbers: linear control of light with light," Nat. Rev. Mater. **2**, 17064 (2017).

3. S. Dutta-Gupta, O. J. F. Martin, S. Dutta Gupta, and G. S. Agarwal, "Controllable coherent perfect absorption in a composite film," Opt. Express **20**, 1330 (2012).

4. Y. D. Chong, L. Ge, H. Cao, and A. D. Stone, "Coherent Perfect Absorbers: Time-Reversed Lasers," Phys. Rev. Lett. **105**, 053901 (2010).

5. A. Mostafazadeh, "Spectral Singularities of Complex Scattering Potentials and Infinite Reflection and Transmission Coefficients at Real Energies," Phys. Rev. Lett. **102**, 220402 (2009).

6. D. Trivedi, A. Madanayake, and A. Krasnok, "Anomalies in light scattering: A circuit-model approach," Phys. Rev. Appl. **22**, 034061 (2024).

7. A. Kodigala, T. Lepetit, Q. Gu, B. Bahari, Y. Fainman, and B. Kanté, "Lasing action from photonic bound states in continuum," Nature **541**, 196–199 (2017).

8. C. W. Hsu, B. Zhen, A. D. Stone, J. D. Joannopoulos, and M. Soljačić, "Bound states in the continuum," Nat. Rev. Mater. **1**, 16048 (2016).

9. S. Kim, A. Krasnok, and A. Alù, "Complex-frequency excitations in photonics and wave physics," Science (1979). **387**, (2025).

10. D. Trivedi, A. Madanayake, and A. Krasnok, "Revealing invisible scattering poles with complex-frequency signals," J. Appl. Phys. **137**, (2025).

11. C. Rasmussen, M. I. N. Rosa, J. Lewton, and M. Ruzzene, "A Lossless Sink Based on Complex Frequency Excitations," Advanced Science **10**, (2023).

12. A. Farhi, A. Mekawy, A. Alù, and D. Stone, "Excitation of absorbing exceptional points in the time domain," Phys. Rev. A (Coll Park). **106**, L031503 (2022).

13. S. Kim, S. Lepeshov, A. Krasnok, and A. Alù, "Beyond Bounds on Light Scattering with Complex Frequency Excitations," Phys. Rev. Lett. **129**, 203601 (2022).

14. F. Binkowski, F. Betz, R. Colom, P. Genevet, and S. Burger, "Poles and zeros in non-Hermitian systems: Application to photonics," Phys. Rev. B **109**, 045414 (2024).

15. R. Ali, "Lighting of a monochromatic scatterer with virtual gain," Phys. Scr. **96**, 095501 (2021).

16. I. Loulas, E.-C. Psychogiou, K. L. Tsakmakidis, and N. Stefanou, "Analytic theory of complex-frequency-aided virtual absorption," Opt. Express **33**, 28333 (2025).

17. G. P. Zouros, I. Loulas, E. Almpanis, A. Krasnok, and K. L. Tsakmakidis, "Anisotropic virtual gain and large tuning of particles' scattering by complex-frequency excitations," Commun. Phys. **7**, 283 (2024).

18. D. G. Baranov, A. Krasnok, and A. Alù, "Coherent virtual absorption based on complex zero excitation for ideal light capturing," Optica **4**, 1457 (2017).

19. D. L. Sounas, "Virtual perfect absorption through modulation of the radiative decay rate," Phys. Rev. B **101**, 104303 (2020).

20. Q. Zhong, L. Simonson, T. Kottos, and R. El-Ganainy, "Coherent virtual absorption of light in microring resonators," Phys. Rev. Res. **2**, 013362 (2020).

21. A. V. Marini, D. Ramaccia, A. Toscano, and F. Bilotti, "Perfect Matching of Reactive Loads Through Complex Frequencies: From Circuital Analysis to Experiments," IEEE Trans. Antennas Propag. **70**, 9641–9651 (2022).

22. G. Trainiti, Y. Ra'di, M. Ruzzene, and A. Alù, "Coherent virtual absorption of elastodynamic waves," Sci. Adv. **5**, (2019).

23. Y.-F. Xia, Z.-X. Xu, Y.-T. Yan, A. Chen, J. Yang, B. Liang, J.-C. Cheng, and J. Christensen, "Observation of Coherent Perfect Acoustic Absorption at an Exceptional Point," Phys. Rev. Lett. **135**, 067001 (2025).

24. F. Ju, C. Liu, Y. Cheng, S. Qian, and X. Liu, "Acoustic coherent perfect absorption based on a PT symmetric coupled Mie resonator system," APL Mater. **11**, (2023).

25. N. Engheta and R. W. Ziolkowski, eds., *Metamaterials* (Wiley, 2006).

26. N. I. Zheludev and Y. S. Kivshar, "From metamaterials to metadevices," Nat. Mater. **11**, 917–924 (2012).

27. F. Monticone, N. M. Estakhri, and A. Alù, "Full Control of Nanoscale Optical Transmission with a Composite Metascreen," Phys. Rev. Lett. **110**, 203903 (2013).

28. Q. Wang, E. T. F. Rogers, B. Gholipour, C.-M. Wang, G. Yuan, J. Teng, and N. I. Zheludev, "Optically reconfigurable metasurfaces and photonic devices based on phase change materials," Nat. Photonics **10**, 60–65 (2016).

29. N. Mohammadi Estakhri and N. Engheta, "Tunable metasurface-based waveplates - A proposal using inverse design," C. R. Phys. **21**, 625–639 (2021).

30. N. M. Estakhri and T. B. Norris, "Tunable quantum two-photon interference with reconfigurable metasurfaces using phase-change materials," Opt. Express **29**, 14245 (2021).

31. Z. Li, R. Pestourie, Z. Lin, S. G. Johnson, and F. Capasso, "Empowering Metasurfaces with Inverse Design: Principles and Applications," ACS Photonics **9**, 2178–2192 (2022).

32. W. Li, H. Barati Sedeh, D. Tsvetkov, W. J. Padilla, S. Ren, J. Malof, and N. M. Litchinitser, "Machine Learning for Engineering Meta-Atoms with Tailored Multipolar Resonances," Laser Photon. Rev. **18**, (2024).

33. D. M. Pozar, *Microwave Engineering*, Fourth edition (John Wiley & Sons, Inc, 2012).

34. P. Bai, K. Ding, G. Wang, J. Luo, Z.-Q. Zhang, C. T. Chan, Y. Wu, and Y. Lai, "Simultaneous realization of a coherent perfect absorber and laser by zero-index media with both gain and loss," Phys. Rev. A (Coll Park). **94**, 063841 (2016).

35. A. Mostafazadeh and M. Sarısaman, "Lasing-threshold condition for oblique TE and TM modes, spectral singularities, and coherent perfect absorption," Phys. Rev. A (Coll Park). **91**, 043804 (2015).

36. K. Zeng, C. Wu, X. Guo, F. Guan, Y. Duan, L. L. Zhang, X. Yang, N. Liu, Q. Dai, and S. Zhang, "Synthesized complex-frequency excitation for ultrasensitive molecular sensing," eLight **4**, 1 (2024).

37. S. Molesky, Z. Lin, A. Y. Piggott, W. Jin, J. Vucković, and A. W. Rodriguez, "Inverse design in nanophotonics," Nat. Photonics **12**, 659–670 (2018).

38. D. Liu, Y. Tan, E. Khoram, and Z. Yu, "Training Deep Neural Networks for the Inverse Design of Nanophotonic Structures," ACS Photonics **5**, 1365–1369 (2018).

39. Z. Liu, D. Zhu, S. P. Rodrigues, K.-T. Lee, and W. Cai, "Generative Model for the Inverse Design of Metasurfaces," Nano Lett. **18**, 6570–6576 (2018).

40. N. Mohammadi Estakhri, B. Edwards, and N. Engheta, "Inverse-designed metastructures that solve equations," Science (1979). **363**, 1333–1338 (2019).

41. A. Vallone, N. M. Estakhri, and N. Mohammadi Estakhri, "Region-specified inverse design of absorption and scattering in nanoparticles by using machine learning," Journal of Physics: Photonics **5**, 024002 (2023).

42. S. Kim, Y.-G. Peng, S. Yves, and A. Alù, "Loss Compensation and Superresolution in Metamaterials with Excitations at Complex Frequencies," Phys. Rev. X **13**, 041024 (2023).

43. A. Krasnok and A. Alu, "Active Nanophotonics," Proceedings of the IEEE **108**, 628–654 (2020).

44. Flexcompute, "Fast, Modern Photonic Simulations: Flexcompute Tidy3D," https://www.flexcompute.com/tidy3d/.